# The Coming Age of Pervasive Data Processing

Jan S. Rellermeyer    Sobhan Omranian Khorasani    Dan Graur    Apourva Parthasarathy

Distributed Systems Group
Delft University of Technology
Delft, Netherlands
{j.s.rellermeyer, s.omraniankhorasani}@tudelft.nl    {d.o.graur, a.parthasarathy}@student.tudelft.nl

*Abstract*—Emerging Big Data analytics and machine learning applications require a significant amount of computational power. While there exists a plethora of large-scale data processing frameworks which thrive in handling the various complexities of data-intensive workloads, the ever-increasing demand of applications have made us reconsider the traditional ways of scaling (e.g., scale-out) and seek new opportunities for improving the performance. In order to prepare for an era where data collection and processing occur on a wide range of devices, from powerful HPC machines to small embedded devices, it is crucial to investigate and eliminate the potential sources of inefficiency in the current state of the art platforms. In this paper, we address the current and upcoming challenges of pervasive data processing and present directions for designing the next generation of large-scale data processing systems.

## I. INTRODUCTION

Many contemporary applications already depend to a large extent on data, either online (i.e., by directly processing the data within the critical path) or offline (e.g., in the form of a trained model derived from data). The proliferation of analytics on big data has resulted in a large ecosystem of solutions for cluster- and datacenter computing that are successfully deployed and deliver important insights to businesses around the world [1]. Modern machine learning is following this trend at an even faster pace with a variety of exciting and demanding applications including self-driving cars [2], processing spoken language fluently [3], or predicting consumer behavior [4]. Both domains have in common that the quality of output tends to depend on the amount of data that is available for processing, which drives the need for continuously scaling up the processing capabilities in order to keep up with the demand. At the same time, the environment and infrastructure for running such applications is getting increasingly complex and distributed. Modern systems often comprise of multiple tiers and span the cloud and mobile or embedded devices in the field. This setup is increasingly complemented by resources on the edge of the network in order to reduce latency [5], [6] or address privacy concerns arising from the centralization of data [7].

The combined result is that data processing is becoming more and more pervasive and embedded into everything, as illustrated by ongoing trends like the Internet of Things [8], smart homes [9], smart manufacturing like Industry 4.0 [10], and mobile edge computing [11]. As we argue in this paper, most of the systems available today are not sufficiently prepared to address the upcoming challenges of pervasive data processing. They typically require excessive manual configuration and tuning in order to get acceptable performance, lack the efficiency required to scale to the growing problem sizes, are unable to leverage the advancements in modern computer architecture, and do not support end-to-end quality of service in complex setups involving multiple tiers.

## II. THE JVM AS THE UNIVERSAL PLATFORM FOR DATA PROCESSING?

Since its humble beginning in 1996 as a runtime for an embedded system inside a handheld home-entertainment device [12] the Java Virtual Machine (JVM) [13] has seen a tremendous boost in adoption and different ecosystems have created themselves around the platform. Examples where the JVM is (still) the gravitational center include application servers [14]–[17] (despite challenged by V8 as the runtime for server-side Javascript such as in Node.js [18]), enterprise middleware [19]–[22], and systems for big data processing. In the latter category, 40 out of 50 big data projects under the Apache umbrella are mainly implemented in a JVM-based language (primarily Java and Scala). The most prominent examples are projects like Hadoop [23], Spark [24], Flink [25], Storm [26], and Kafka [27].

Two of its main design features have propelled the JVM to its current state of adoption: automatic memory management through a garbage-collected heap and the *Write once, run everywhere* paradigm by using platform-independent bytecode as an intermediate representation and relying on a mixture of interpretation and just-in-time compilation (JIT) around hotspots for execution.

While the first feature is primarily associated with an increase of developer productivity and runtime-safety by reducing the chances of unintentional memory leaks, the second feature has mostly benefited applications intended to run on a large variety of different machines which then becomes trivial due to the virtual machine abstraction. Unfortunately, in practice both features do not come without consequences.

Garbage collection is widely recognized as a potential performance bottleneck and negatively affecting latency predictability [28], even though the precise level of impact is a function of the workload and the collection algorithm used [29]. Furthermore, the setup of big data processing invalidates the weak-generational hypothesis which provides the justification for the now predominantly used approach of generational garbage collection: The vast majority of the objects, i.e., the data objects, live for the majority of the duration of the program and therefore never become collectible. Various domain-specific solutions have been proposed (e.g., [30]–[32]) but a growing group of big data processing platforms have abandoned the idea of using the garbage-collected heap for the *data objects* itself and instead moved them to unmanaged off-heap memory [33]. This also addresses a second problem with the managed heap. Objects in the JVM are not stored in a contiguous manner but can be spread across the entire heap and not even arrays are guaranteed to always be contiguous. This makes it challenging for any form of accelerator hardware like GPUs or FPGAs to handle the data since those devices typically require it to be transferred into device memory. In the absence of a compact layout, this inevitably results in excessive pointer chasing and intermediate copying.

This leaves platform-independence as the primary contribution of the JVM for big data processing. Unfortunately, again the disadvantages can outweigh the advantages as exemplified by the impact of the JVM startup time on the job execution time [34]. This overhead comprises of the initialization time of the VM itself and the classloading time that is required to bring the application into a runnable state. While in the original application domain of Java, the platform independence proved to be a great asset, modern datacenters exhibit only a very limited degree of heterogeneity. Furthermore, modern big data platforms dwarf the size of the user programs that implement the actual application. As a result, the JVM instantiates mostly the same identical classpath over and over, which makes the approach of always starting *cold* from raw bytecode questionable. This is particularly a problem when the task length is small, which is common in big data processing pipelines. Studies on characteristics of MapReduce based cluster workloads show that majority of jobs are short-lived and follow a heavy tailed distribution [35], [36]. An analysis of production traces from a Hadoop cluster has shown that about 80% of jobs in the cluster are small, running for less than 4 minutes and 90% of jobs read less that 64MB of data from HDFS [36].

The issue of classloader overhead was partly addressed by the introduction of Class Data Sharing (*CDS*), a technique to persist the metadata of classes loaded during the execution of an application in a memory mapped file so that any subsequent run of the application can simply map this file into memory and reuse the loaded classes. This feature was introduced in JDK 1.5 [37] but until recently was only able to archive classes loaded by the Bootstrap class, e.g., JRE classes from the `java.lang` package. Support for archiving application classes was added to HotSpot with Java 9 and became part of OpenJDK with release 10. As an additional benefit to reducing classloading overhead, the CDS archives become sharable among multiple running JVM instances, which reduces the memory overhead of the classpath when running multiple instances of the JVM with the same classpath. At that point, CDS is essentially becoming equivalent to shared libraries for the operating system's dynamic loader.

| Workload | Job Execution time (s) with CL | Job Execution time (s) with CDS | Improvement |
|---|---|---|---|
| Hadoop wc Tiny | 49.73 | 36.04 | 27.5% |
| Hadoop wc Small | 118.49 | 100.28 | 15.3% |
| Hadoop wc Large | 779.17 | 713.92 | 8.4% |

TABLE I: Reduction in Job Execution Time in Hadoop WordCount through CDS

We conducted preliminary experiments running WordCount on Hadoop in three different settings, with a *Tiny* workload of only 32kiB of input data, with a *Small* data set of 320MiB, and with the *Large* set of 3.20GiB, all three from the HiBench [38] benchmark suite. We assess the classloading overhead of Hadoop when executed with traditional classloading versus with a CDS archive containing the content of the entire Hadoop application classpath. We use Apache Hadoop version 3.1.0 on OpenJDK 10.0 which has full application classpath CDS support. Hadoop currently does not run on OpenJDK9+ out of the box but requires the use of several patches to make it compatible. We applied patches[1] 12760, 14984, 14986, 15133, 15304, 15775, and 15610. The data is stored on HDFS and all workloads are managed by the YARN resource manager. The experiments were conducted on a single compute node on the DAS5 [39] cluster with 64GiB main memory and a dual 8 core Intel Xeon processor. Table I shows the reduction in execution time for the three workloads. As expected, the impact is the highest for short running jobs with about 27.5% reduction for the *Tiny* data set. However, even with the *Large* data set the improvement is still 8.4%. In addition to the reduction of job execution time, CDS also provides a benefit in terms of memory consumption in multi-tenant setups where multiple instances of Hadoop run on the same machine simultaneously.

While the support for CDS archives was steadily improved over the past releases of the JVM, until now the JVM only permitted the loading of a single CDS archive at the time. In recent work, however, we have extended OpenJDK to support an arbitrary number of CDS archives which allows for full modularity of applications. Figure 1 shows a simplified version of the modular CDS in which instead of having one large monolithic CDS archive that includes every class, each framework (e.g., Spark) gets its own separate archive which itself consists of smaller set of archives (e.g., HDFS, JDK) to provide full modularization as well as layered CDS. We see this as an important step towards more sharing between JVM instances and a reduction of the startup overhead, issues that are increasingly becoming a problem for large-scale applications in cluster and cloud computing.

HotTub [34] tried to address the same challenges by making the JVM *reusable*. OpenJDK is modified in such a way that after a program invocation the state of the JVM is largely restored to right after the initialization of the system so that future invocations of the same application do not incur the startup overhead and can benefit from a warmed-up classpath.

---

[1] https://issues.apache.org/jira/projects/HADOOP/issues

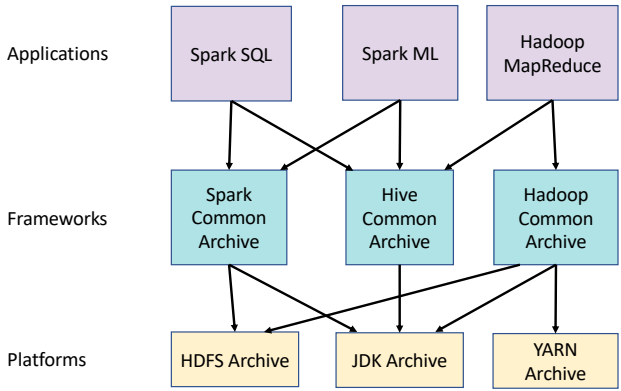

Fig. 1: Modular Class Data Sharing for Big Data Processing Platforms

While the approach is certainly interesting, our attempt to run the system in a production environment resulted in the entire machine crashing after merely half an hour. An in-depth analysis revealed a subtle but important problem. In a cluster environment, applications do not necessarily get the chance to gracefully shut down. In fact, cluster scheduler like YARN are aggressively terminating workers by sending signals to the JVM. However, this challenges the HotTub JVM to take action and bring down any remaining threads in order to restore a clean state. After attempting a cooperative shutdown by interrupting threads, the mechanism unfortunately falls back to using `ThreadDeath`, the facility behind `Thread.stop()`, a call that has been deprecated since the release of Java version 1.2 in 1998 since it is inherently unsafe and can lead to deadlocks [40]. To make matters worse, there are even plenty of reasons why stopping the thread can be unsuccessful, e.g., a thread being blocked by an `epoll` or similar system calls. However, when the thread survives it continues to hold memory on and off the heap as well as file descriptors. This inevitably leads to an accumulating leak of system resources in practice which ultimately overwhelms the machine.

Attempting to forcibly terminate individual threads with the goal of letting the process survive is–independently of the JVM–fundamentally unsound and the same problem would occur at the level of pthreads when using `pthread_cancel` while relying on the thread having had the foresight to register reasonable `pthread_cleanup_push` handlers for resource cleanup. As a matter of fact, the majority of the code in the wild relies at least partially on the operating system to clean up systems resources when the process terminates and delaying the process termination for the purpose of reuse is breaking this assumption.

After our experience with HotTub, we concluded that reusing the JVM process is not the right approach. Instead, we envision a JVM design that makes the system *forkable* at an early stage to avoid the startup overhead while increasing the degree of sharing between child JVMs to create synergies between the instances, which we have successfully achieved, e.g., with our enhanced modular CDS and in prior work that aimed at making significant structural changes towards turning the JVM from a single-application into a multi-tenant platform [41].

More fundamentally, however, it can be questioned if the JVM is generally the right platform for big data processing in light of the disadvantages of its design potentially erasing its advantages. In a second line of research, we are joining a growing list of efforts (e.g., [42], [43]) to implement big data processing clos(er) to the machine and closing the performance gap between big data and HPC.

### III. FROM SCALE OUT TO HPC? THE PROBLEM OF SUSTAINABILITY

With the growing volumes of data and the increasing demand for deriving insights out of them, the systems need to become more efficient to avoid an equally increasing carbon footprint of big data processing. This, however, means that the currently predominant approach of *scaling out* is not viable in the long run since adding more machines to outnumber the problem leads to a proportionally equivalent increase in energy consumption. High performance computing faced similar challenges that only adding more CPUs and memory alone is insufficient to satisfy the ever-increasing appetite for running even more demanding simulations and applications on supercomputer machinery. A great amount of attention has been dedicated to the problem of increasing the utilization of the existing machinery [42], [43] and to increase the performance per watt by introducing specialized hardware in the form of accelerators [44], [45].

With the slowdown and potentially demise of Moore's Law [46], new generations of computer hardware are no longer able to provide additional performance *for free*, i.e., in a completely transparent manner. Instead, recent improvements have led to a more complex computer architecture with the introduction of multi-core CPUs and non-uniform memory (NUMA). As a result, programmers now have to structure their code accordingly to leverage the advantages of the next generation processors.

In recent work, we conducted an experiment to see how the NUMA architecture affects Spark. We ran the Terasort workload in three different settings on a two-socket Intel server with 32GiB of memory per socket. In the default setting, memory placement is opportunistic and follows the current rules in Linux that memory is primarily allocated on the NUMA node on which the thread ran that requested the memory. In the two other settings, we ran two separate instances of Spark both pinned to one socket and also forced the memory allocations to remain either on the same NUMA node for best locality, or, as a worst-case scenario, always happen on the remote node. Table II summarizes the results and shows in the worst case, when all memory accesses are remote then this can cost up to 53.8% of performance.

| # nodes | runtime(s) | | |
|---|---|---|---|
| | **remote** | default | **local** |
| 1 | 377 (+53.8%) | 245 (0%) | 205 (-16.3%) |
| 2 | 198 (+46.6%) | 135 (0%) | 111 (-17.7%) |
| 3 | 132 (+33.3%) | 99 (0%) | 83 (-16.1%) |
| 4 | 111 (+32.1%) | 84 (0%) | 73 (-13%) |

TABLE II: NUMA effect on the runtime of Terasort using Spark

With the simple structural change of not treating the two-socket machine as one coherent system but rather as two independent systems and consequently running two separate Spark instances on them, the performance is improved by up to 18%.

Memory is indeed just one aspect in which modern computer architecture deviates from more uniform traditional systems. The use of GPUs and FPGAs as accelerators makes compute increasingly more heterogeneous. Traditional interfaces such as the Unix IO Subsystem are increasingly bypassed and extended to address the properties of modern storage devices such as in the case of Open Channel SSDs [47]. Even the network is becoming a more active part in systems design through the introduction of software-defined networking [48] and offloading techniques such as RDMA [49].

Although this indicates how respecting the properties of the underlying hardware can have a significant impact, the virtualized environments such as the JVM in which big data frameworks run often fail to make the best use of these characteristics due to their abstract and high-level nature. Even the use of a high-level programming language scan can have an impact. Recent research suggests significant differences in the resulting energy efficiency and runtime [50]. This again raises the question whether using environments such as the JVM, which deliberately abstracts away the details of the underlying architecture, makes us miss optimization opportunities that we are going to need in order to keep up with the demand for tomorrow's pervasive data processing applications in a sustainable way.

IV. EMERGING WORKLOADS - WILL BIG DATA PROCESSING AND MACHINE LEARNING EVENTUALLY CONVERGE?

Machine learning techniques have seen a tremendous increase of adoption over the last decade and this growth has been fueled by both advances on the algorithmic level and the ability to leverage hardware acceleration for making the algorithms run at scale. However, in order to increase the quality of predictions and make machine learning applicable to more complex applications, large amounts of training data need to be processed. Since this can easily exceed the capabilities of a single machine, purposely distributed machine learning systems like Google's TensorFlow [51], Microsoft's CNTK [52], or Facebook's Caffe2 [53] have become increasingly popular while originally single-machine libraries like Theano [54] and Keras [55] added support for distribution by essentially transforming themselves into programming abstractions and leveraging these systems for distributed execution.

It is not unreasonable to assume that there will be an even stronger convergence in the future between big data processing and machine learning. On the big data side, classic big data platforms have already added explicit support for machine learning algorithms, e.g., Hadoop through Mahout [56] and Spark through MLlib [57]. Besides trying to appeal to users of these emerging workloads and easing the integration of machine learning into more complex data processing pipelines, there is also a noticeable shift in the industry from a largely reactive model of analytics–which is well supported by the state-of-the-art batch processing architecture and map-reduce like programming models–towards an increasingly predictive model. The latter requires continuous processing of data more akin to training a model and makes probabilistic methods more appealing.

On the machine learning side, the desire to achieve better accuracy and to be able to apply machine learning model to more complex problems leads to increasing model sizes which in turn requires the processing of much larger volumes of training data. Therefore, the efficient parallel processing of the data is increasingly becoming important, which steers solution developers to using distributed systems. However, in such setups the architecture of the machine learning platform has tremendous impact on the resulting performance.

Figure 2 shows the CPU utilization and runtime of an experiment running ResNet50 on TensorFlow[2]. For the same problem on the same system running on the same hardware (nodes in the DAS-5 cluster [39]), the runtime varies between just under 400 seconds for a Collective AllReduce architecture (Figure 2c) to almost 700 seconds for a co-located Parameter Server (Figure 2a).

Interestingly, problems like ResNet also show clear patterns of also being very sensitive to the available network bandwidth, which differentiates them from classic big data processing for which Ousterhout et al. claim that the network does not have an influence on performance [58]. This makes it a potential candidate for advanced networking methods as well as the use of high-bandwidth interconnects.

V. AN OPPORTUNITY FOR SYSTEMS RESEARCH

Both the coming generation of big data processing and distributed machine learning systems are poised to increase efficiency by reducing overhead in execution imposed by the runtime system and embracing accelerators as well as other methodologies from high-performance computing. From the perspective of distributed systems, the problem of placing and moving data at maximum efficiency is becoming the most important problem to solve since the computation is increasingly offloaded to specialized hardware such as GPUs or FPGAs.

---

[2]Data: Synthetic ImageNet, initially 300x300x3, then cropped to 224x224x3. Per-device batch size: 128 images / batch, 32-bit floats: yes, Optimizer: Stochastic Gradient Descent, Initial Learning Rate: 0.1, Forward and Backward: yes, Warmup Runs: 5, Evaluation Runs: 10, each run is a batch

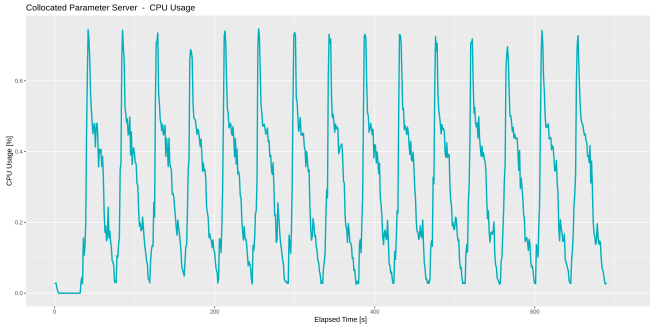

(a) Parameter Server - Co-located

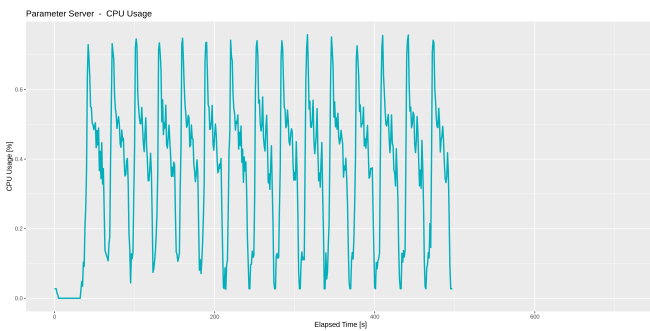

(b) Parameter Server - Separate Machine

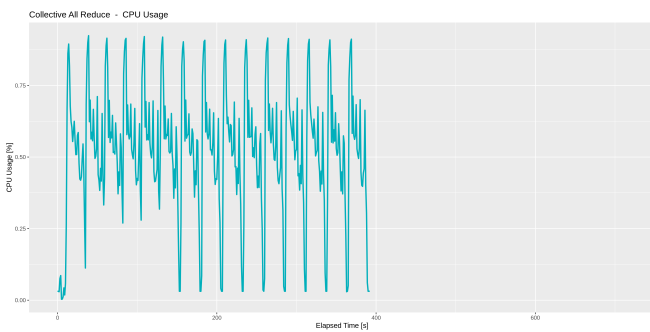

(c) Collective AllReduce

Fig. 2: ResNet50 on TensorFlow using 9 nodes in different communication patterns and topologies

At the same time, the network is shaping itself into a much more active part of the system as exemplified by Software-Defined Networking (SDN) or Remote Direct Memory Access (RDMA). The upcoming trends in micro-architectures promise to go beyond NUMA and introduce additional heterogeneity into the system through technology like 3D-stacked logic-in-memory modules [59] which are capable of running a limited amount of processing directly on the memory.

In the history of data processing, prior technology reached similar critical points of transition between an era dominated by the quest of the proper functionality and algebraic model and the following era characterized by making the technology run efficiently on what was back then modern hardware and at sufficient scale to make it ready for wide adoption in the enterprise world. This effort, however, partly conflicted with the role of the operating system as the central authority for making scheduling decisions and multiplexing resources among different processes. Ultimately, some of the most successful commercial database system ended up bypassing the OS for critical functionality like organizing the layout of the data on disk for the sake of additional performance. It is interesting to see if big data processing or distributed machine learning will reach a similarly essential role for enterprises that designers are willing to sacrifice generality for additional performance. The tension between the capabilities of modern and emerging hardware on the one side, the requirements of novel large-scale systems on the other side, and the available APIs in between that still largely resemble the world of the 1980s as in the case of POSIX is building up. It might at some point become high enough to overcome the gravitational force of backwards-compatibility and compliance with traditional standards and open the door for new generations of systems.

The historic comparison with classic databases, however, also exposes an important perceived weakness of today's large scale data processing system: The strong need for manual tuning in order to get peak performance. Relational database systems operate on a declarative query interface and an algebraic model of queries and data which allows the engine to optimize query plans based on a combination of algebraic transformation and heuristics. Systems like Spark for big data processing or TensorFlow for distributed machine learning do not come even close to the same level of usability and instead require the operator to understand and imperatively specify fundamental properties of the system (e.g., the choice of algorithm and partitioning for generic operations like joining two data sets) while then still requiring an excessive amount of fine-tuning of configuration parameter that potentially impact the resulting performance of the user program. Our TensorFlow experiment (Figure 2) greatly illustrates this issue since the default communication model—parameter server—performs significantly worse than AllReduce which the user has to manually select. This can indeed become a serious burden to a more widespread adoption.

In order to tackle the challenges, we postulate the following principles for system software for next generation big data processing and distributed ML systems:

- **Respect the hardware**: Efficiency is key to increasing the power of both big data and ML. Building systems that are agnostic to the hardware they run on is no longer sustainable.
- **Co-Design** the operating system and the data processing system. Efficient management of the flow of data and the ability to implement and enforce end-to-end QOS requires a richer interface between the system (or the middleware) and the OS than traditional POSIX.
- **Go with the Flow**: The real challenge of high-performance data processing is to efficiently manage the flow of data between systems and components.
- **Make systems smarter** in their ability to perform automatic performance tuning and selecting the optimal

combination of algorithm and topology based on features derived from the user program.

We see each of these principles as essential for taking big data processing to the next level: large scale and deployed virtually everywhere. The term *scale* is not meant to be restricted to the volume of data but indeed encompass all four V's of big data processing [60]. In order to reach these new dimensions of scale in a sustainable way, we need novel systems that leverage the upcoming improvements of the hardware which are no longer coming "for free" as in the golden era of Moore's Law but instead require adjustments to take advantage of an increasing architectural diversity.

## VI. RELATED WORK

With the rapid growth in the size of the workloads, optimizing the performance of big data processing platforms has been the focus of many researches. Some of the existing approaches in combining HPC and big data include Nimbus [42] and Thrill [43] in which performance increase is achieved through producing highly-optimized native code. Moreover, there have been efforts in gaining more efficiency out of the existing infrastructure by leveraging hardware accelerators such as GPUs and FPGAs. For instance, Microsoft's project brainwave [45] leverages FPGAs to accelerate real time AI calculations. In [44], an FPGA-based Spark implementation is presented which acquires 1.79x speedup compared to the CPU implementation. Spark-GPU [61] accelerates Spark workloads through GPUs and reports a speedup of 16.13x for machine learning workloads and 4.83x for SQL queries.

Other works have focused on putting the computer architecture itself to use in order to improve the performance. In [62], authors present an RDMA-based Spark on InfiniBand-enabled clusters in which the performance of SQL and graph processing workloads are improved by 32% and 46% respectively. As discussed in the previous section, these efforts emphasize the importance of respecting the underlying features of the hardware.

In machine learning, the use of GPUs as accelerators is by now best practice. An alternative to generic GPUs for acceleration is the use of Application Specific Integrated Circuits (ASICs) which implement specialized functions in a highly optimized design. The demand for such chips has increased significantly in the past years [63]. Google, for instance, developed their Tensor Processing Unit (TPU) [64], which is an ASIC designed to accelerate their TensorFlow platform. The benefit of the TPU over regular CPU/GPU setups is not only its increased processing power but also its power efficiency, which is important in large-scale applications due to the cost of energy and limited availability in large-scale data centers. In experiments, the TPU approached a 200x improvement of the performance per watt compared to a commodity CPU [65]. Further benchmarking indicated that the total processing power of a TPU or GPU can be up to 70x higher than a CPU for a typical neural network [64].

## VII. CONCLUSIONS

There are several technological pushes on the hardware side ranging from micro-architecture over peripheral devices to networks which all promise key performance improvements when leveraged accordingly. In this article, we have outlined some of the challenges that current systems experience when trying to actually do so. With more heterogeneous machines and novel technology on the horizon, this divide is only going to become wider. In some of our prior work, we have attempted to address some of the friction between existing software systems, platforms, and new technology in terms of building distributed systems for the Internet of Things [66], increasing scalability and performance [67], [68], and dealing better with multi-tenancy [41]. However, we believe that at this point it is time to rethink the entire stack and co-design new interfaces between hardware, operating system, and data processing systems. While this constitutes a significant effort and requires a broad view of all the layers involved, we see it as inevitable in order to build novel applications in a more sustainable and efficient way for the coming age of pervasive data processing.